\documentclass[twocolumn,twoside,slac_two]{revtex4}
\usepackage{graphicx}
\usepackage{fancyhdr}
\begin{document}

\title{Tuning into UHE Neutrinos in Antarctica - The ANITA Experiment}


\author{P.~Mio\v{c}inovi\'c, for the ANITA Collaboration}
\affiliation{University of Hawaii, Dept. of Physics and Astronomy, Honolulu, HI 96822, USA}
\author{The ANITA Collaboration:\\
S.~W.~Barwick$^{3}$,
J.~J.~Beatty$^{7}$, 
D.~Z.~Besson$^{6}$, 
W.~R.~Binns$^{8}$,
B.~Cai$^{10}$,
J.~M.~Clem$^{1}$,
A.~Connolly$^{4}$,
S.~Coutu$^{2}$,
D.~F.~Cowen$^{2}$,
P.~F.~Dowkontt$^{8}$,
M.~A.~DuVernois$^{10}$, 
P.~A.~Evenson$^{1}$, 
D.~Goldstein$^{3}$,
P.~W.~Gorham$^{5}$, 
C.~L.~Hebert$^{5}$,
M.~H.~Israel$^{8}$, 
J.~G.~Learned$^{5}$, 
K.~M.~Liewer$^{9}$,
J.~T.~Link$^{5}$,
S.~Matsuno$^{5}$,
P.~Mio\v{c}inovi\'c$^{5}$,
J.~W.~Nam$^{3}$,
C.~J.~Naudet$^{9}$,
R.~Nichol$^{7}$,
K.~J.~Palladino$^{7}$,
M.~Rosen$^{5}$,
D.~Saltzberg$^{4}$, 
D.~Seckel$^{1}$,
A.~Silvestri$^{3}$,
G.~S.~Varner$^{5}$,
D.~Williams$^{4}$
}
\affiliation{
(1) Bartol Research Institute, University of Delaware, Newark, DE 19716, USA
\newline
(2) Dept. of Astronomy and Astrophysics, Penn. State University, University Park, PA 16802, USA 
\newline
(3) Dept. of Physics and Astronomy, University of California, Irvine CA 92697, USA 
\newline
(4) Dept. of Physics and Astronomy, University of California, Los Angeles, CA 90095, USA
\newline
(5) Dept. of Physics and Astronomy, University of Hawaii, Manoa, HI 96822, USA
\newline
(6) Dept. of Physics and Astronomy, University of Kansas, Lawrence, KS 66045, USA
\newline
(7) Dept. of Physics, Ohio State University, Columbus, OH 43210, USA
\newline
(8) Dept. of Physics, Washington University in St. Louis, MO 63130, USA
\newline
(9) Jet Propulsion Laboratory, Pasadena, CA 91109, USA
\newline
(10) School of Physics and Astronomy, University of Minnesota, Minneapolis, MN 55455, USA
}

\begin{abstract}
The Antarctic Impulsive Transient Antenna (ANITA) experiment is being developed to search for ultra-high-energy (UHE) neutrino interactions ($>3\times10^{18}$ eV) in the Antarctic ice cap. A neutrino interaction in the ice will produce a radio pulse by the means of the Askaryan effect. The large radio transparency of ice allows for such a pulse to be recorded by a cluster of balloon-borne antennas. The details of the ANITA instrument, now in a construction phase, and the science we hope to achieve is discussed.
In order to prepare for the main mission, we have flown ANITA-lite during the 2003/04 austral season. ANITA-lite consisted of two quad-ridge horn antennas and a prototype RF (radio frequency) triggering and recording system. Here we present the results of an impulsive RF background survey of Antarctica, as well as proof-of-principle gain, tracking, and timing calibrations conducted by observing solar radio emissions and calibration radio-pulses. A preliminary UHE neutrino flux limit based on ANITA-lite data is also presented.
\end{abstract}

\maketitle

\thispagestyle{fancy}


\section{INTRODUCTION}

High-energy neutrinos ($>$few tens of GeV) are currently an unexplored information channel available to particle 
astrophysicists. They are the third leg necessary to support current astrophysical theories, and 
combined with optical and cosmic-ray observations can provide a more clear view into the inner 
workings of the most energetic engines in the Universe. Due to their low interaction cross-sections, 
neutrinos are devilishly 
hard to work with, and only in the last decade experimental physics has been able to provide first
definitive results based on large scale neutrino observations. At the same time, the power law nature of
astrophysical accelerators means that a particle flux decreases rapidly as one looks at higher energies.
These two facts imply that for high-energy neutrino observations one needs to instrument a very large 
target volume; for the expected fluxes at TeV energies observatories of $\sim$1 km$^3\cdot$sr are 
required, while at EeV energies one needs $\sim$1000 km$^3\cdot$sr. 
While neutrino observatories at 1 km$^3$ scale are currently being constructed~\cite{icecube}, 
{\it in-situ} instrumentation of much larger volumes is impractical. However, substantial
transparency of cold ice ($<-20^\circ$C) to radio frequency radiation, combined with enhancement of
 radio emissions from 
high-energy neutrino-induced showers in ice due to the Askaryan effect, offers a promising way for 
reaching the required target volumes. A balloon-borne radio telescope at an altitude of 35-40 km 
above Antarctica would observe $\sim10^6$ km$^3$ of radio-transparent ice. In this report, 
we describe plans for such an experiment, starting with a summary of Askaryan effect physics
and the science goals of the mission. We also present results from the prototype flight and 
describe expected sensitivity to ultra-high-energy neutrino flux.    
      
\subsection{Askaryan effect}

G.~Askaryan proposed in 1962~\cite{askaryan} that a compact particle shower will produce a coherent 
radio Cherenkov emission. Subsequent theoretical work in the 80's~\cite{zhelez1,zhelez2,zhelez3} and 
the 90's~\cite{zhs} supported this prediction. The experimental verification came in 
2001~\cite{slac01}, with follow up measurements confirming frequency and polarization properties of 
the emitted radiation~\cite{slac05}. 

The emission of coherent radio signal comes about from an appearance of the charge asymmetry as a 
particle shower develops in a dense medium. This asymmetry is due to combined effects of positron 
annihilation and Compton scattering of electrons at rest. There is $\sim$20\% excess of
electrons over positrons in such a particle cascade, which moves as a compact bunch a few cm wide and $\sim$1
cm thick at the velocity above the speed of light in the medium. The frequency dependence of 
Cherenkov radiation emitted is $dP\propto\nu d\nu$. In addition, for radiation with wavelength
 $\lambda\gg l$, where $l$ is the scale of the particle bunch, the radiated signal will add coherently and thus
be proportional to the square of shower energy. 

A radio signal emitted by a particle shower in a material such as ice is coherent up to few GHz,
is linearly polarized, and lasts only about a nanosecond. A neutrino with energy of $10^{19}$ eV 
interacting in the ice produces a radio pulse with a peak strength of $\sim10^{-3}$~V/m/MHz at 
a distance of 1~km. 

Several experiments have already utilized Askaryan effect to search for high-energy neutrino interactions, 
RICE at the South Pole~\cite{rice}, FORTE in Greenland ice cap~\cite{forte}, and GLUE in the lunar
regolith~\cite{glue}.
  
\subsection{Science Goals}

Unlike photons, high-energy astrophysical neutrinos propagate through the Universe unattenuated. They
carry information from distances beyond the photon horizon which extends only to few hundred Mpc in 
10-100 TeV range, and even less at PeV energies. Thus, they stand as a unique probe to acceleration
processes associated with the sources of the highest energy cosmic-rays which extend to $10^8$ TeV. 
Additionally, the GZK process~\cite{greisen,zk}, interaction of high-energy cosmic-ray protons with 
the cosmic microwave background, produces high-energy neutrino flux. Observation or strict constraints on 
such a neutrino flux is crucial to resolution of GZK cutoff, which is currently one of the most
controversial topics in cosmic ray physics. 

Beyond the Standard Model, certain models of microscopic hidden dimensions predict that high-energy 
neutrino interactions in ice
could produce highly unstable micro black holes (MBH)~\cite{feng,alvarez_bh}. The decay of these MBH
via Hawking radiation would produce hadronic showers detectable through the Askaryan effect. The 
signature of this process would be an increase in the expected event rate with a strong energy
dependency. Besides observation of extra dimensions, any top-down model of high-energy cosmic-ray 
production produces an associated neutrino flux~\cite{barbot}, which would also be detectable. 

\begin{figure}
\includegraphics[width=75mm]{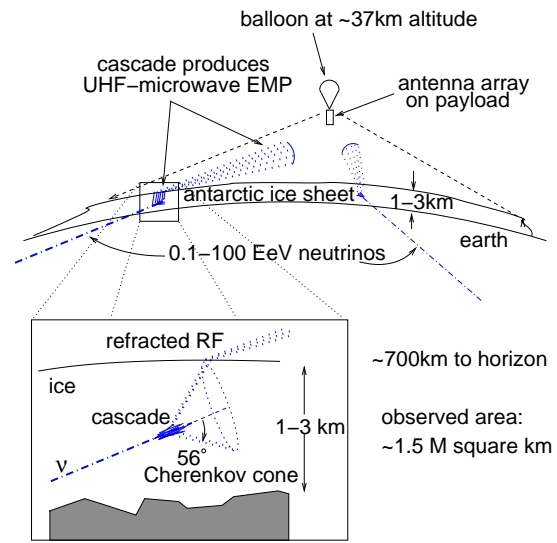}
\caption{\label{fig:detection}A schematic of ANITA detection concept.}
\end{figure}
\section{ANITA}

\subsection{The Detection Concept}

The ANtarctic Impulsive Transient Array (ANITA) has been designed to detect radio pulses emitted by 
neutrino interactions in Antarctic ice sheet. At energies above $10^{18}$ eV, the Earth is opaque to 
neutrinos, so ANITA would be sensitive only to neutrinos arriving at glancing angle with 
respect to the ice surface (Figure~\ref{fig:detection}). 
Neutrinos interacting in the ice will produce a radio pulse emitted along
a Cherenkov cone. This pulse will suffer very little attenuation 
before refracting at the surface and exiting the ice~\cite{barwick_ice}.  The interaction point and the
direction of an incoming neutrino is determined by the time difference in radio pulse arrival between 
antennas and by the polarization of the pulse. 

\begin{figure}
\includegraphics[width=75mm]{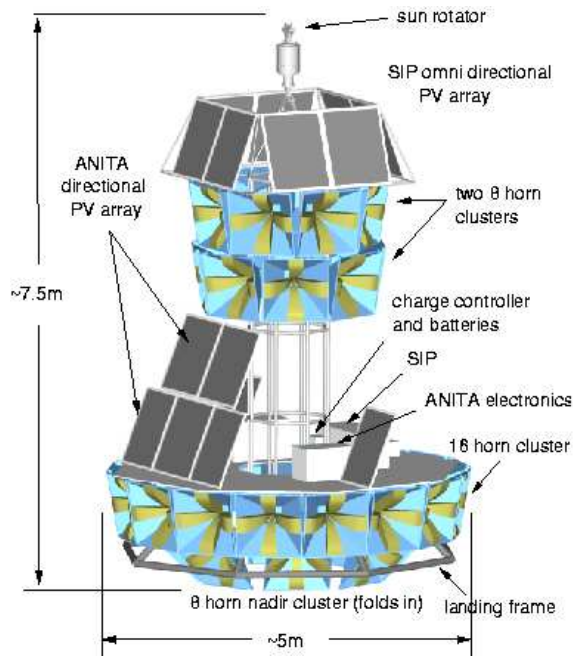}
\caption{\label{fig:anita}A mechanical model of the ANITA instrument.}
\end{figure}
\subsection{The Detector}

The ANITA instrument (Figure~\ref{fig:anita}) will consist of 40 dual polarization, quad-ridged
horn antennas arranged in cylindrically symmetric upper and lower clusters. Each antenna records two 
linear polarizations 
of an incoming radio pulse and has a beam width of about $60^\circ$. These antennas operate over 0.2 to 1.2 
GHz, have approximately 10 dBi gain over the entire frequency range, and a very small phase dispersion,
resulting in a sub-nanosecond impulse response.   
The separation between antenna clusters provides for a vertical timing baseline needed to determine
the elevation angle of the arriving pulse. An overlap between adjacent antenna beam patterns provides 
a pulse gradiometry for determination of the azimuthal angle. The absolute azimuthal orientation will be
measured by Sun sensor instruments, while the instrument tilt will be monitored by a differential GPS 
unit. 
The instrument will trigger on a coincident increase in RF power in several antenna channels. On 
trigger, all antennas will be readout and digitized at $\sim$3GSa/s sampling rate, allowing a 
capture of $\sim$85 ns of data per trigger. The DAQ system will be able to record data rates up to 
5~Hz.    

\section{ANITA-lite}

ANITA-lite was a two antenna prototype of ANITA that was flown during the 2003/04 Antarctic season as a 
piggyback instrument on board Trans-Iron Galactic Element Recorder (TIGER) payload~\cite{tiger}. The goals
of ANITA-lite were to conduct an RF background survey of Antarctica and to test prototype antennas and 
RF electronics which are to be used for ANITA. We have also performed the timing calibration of 
the system by sending RF pulses from the ground station to the instrument in flight.    
\begin{figure}
\includegraphics[width=75mm]{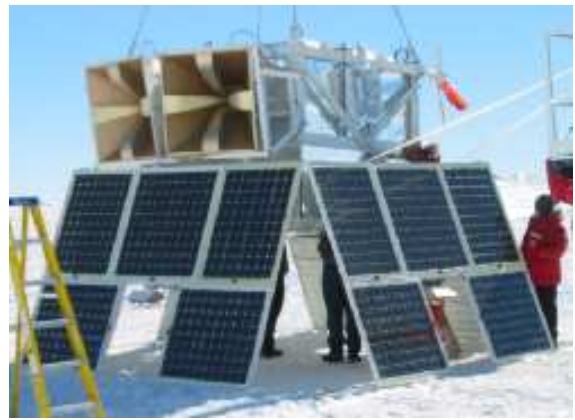}
\caption{\label{fig:anitalite}The TIGER/ANITA-lite payload prior to launch.}
\end{figure}

The ANITA-lite instrument triggered on a coincident increase in RF power in both antennas. The data were 
digitized at 
2~GSa/s, with 512~ns of data per trigger. We collected the total of 130,000 triggers (including 
the data taking and the calibration) over 18~days. 

\subsection{Timing calibration}

\begin{figure}
\includegraphics[width=65mm]{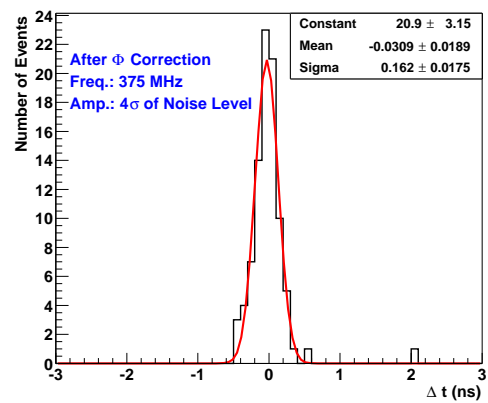}
\caption{\label{fig:timecal}Distribution of time differences between co-polarized channels.}
\end{figure}

Tone bursts at various frequencies and of various durations were sent with a ridged horn antenna from 
the ground to ANITA-lite at float. The signals were observed out to distances greater than 200 km, the main 
limitation being the loss of line-of-sight due to the Trans-antarctic mountain range. Calibration 
signals were bandpass filtered around the calibration frequencies, and a time reference for each antenna
was measured by interpolating the zero-crossing of a signal. Correcting for the separation between
the two antennas, we measure the uncertainty in the signal arrival time between co-polarized 
channels of $\sigma_{\Delta t}=0.16$~ns 
(Figure~\ref{fig:timecal}). 

Based on this measurement, we estimate the time resolution between antenna clusters in ANITA to improve to
$\sigma_{\Delta t}=0.1$ ns due to the increase in number of channels receiving the signal and higher
sampling speed. Since the error
in the pulse arrival angle for antennas separated by the distance $d$ is approximately given by
\begin{equation}
\sigma_\theta\approx \frac{180}{\pi}\frac{\sigma_{\Delta t} c}{d},
\end{equation}
we estimate $0.5^\circ$ and $1.5^\circ$ uncertainty in determining the elevation and azimuth 
angles of arrival, respectively, for pulses originating near the horizon.  

\subsection{Thermal background survey}

\begin{figure}
\includegraphics[width=65mm]{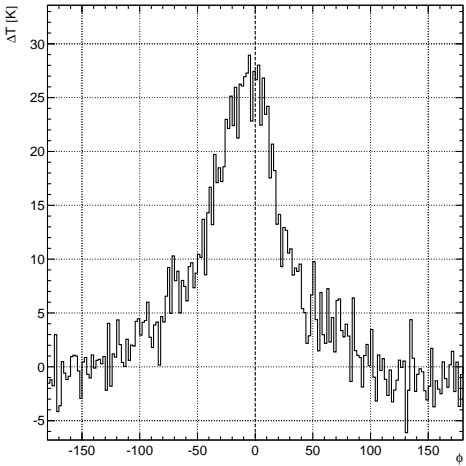}
\caption{\label{fig:asy}The excess effective antenna temperature as a function of azimuth between
the direction of the Galactic Center and the ANITA-lite orientation. The asymmetry in the 
distribution is due to location of the Sun relative to the Galactic Center.}
\end{figure}

In addition to observing the Antarctic ice sheet, ANITA-lite antennas have a significant portion of the 
sky in their field-of-view. As the instrument rotates, the two brightest radio sources in the sky, the Sun and
the Galactic Center (GC), come into view (Figure~\ref{fig:asy}).
\begin{figure}
\includegraphics[width=75mm]{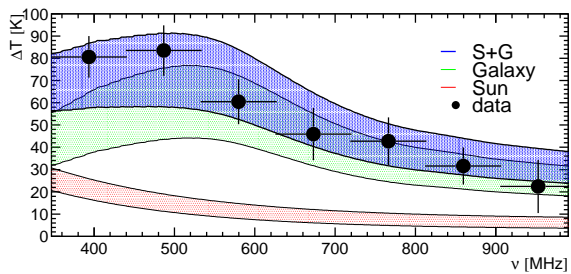}
\caption{\label{fig:ycal}An average increase in the effective antenna temperature when pointing to 
the Sun and the Galactic Center. The top band is a model of the expected temperature increase, the width 
of which accounts for systematic uncertainties. The lower two bands give contributions due to galactic
and solar emissions, respectively. The antenna frequency response is folded into the model.}
\end{figure}
During the ANITA-lite campaign, they were separated by 5-14 
degrees, making it impossible to resolve them with our broad-beam antennas. Figure~\ref{fig:ycal}
compares an average increase in the observed background radiation, expressed in terms of an effective 
antenna temperature, as a function of frequency when pointed in the direction of the Sun and the GC 
with a model of the expected temperature increase. The model was based on the quiet Sun 
model~\cite{sun} corrected for the observed solar activity during the ANITA-lite flight~\cite{suncorr}
and on the all-sky radio survey at 408 MHz~\cite{haslam} with an assumption of synchrotron-like 
frequency dependency of radio emissions from the galactic plane. 
The good agreement between measured and expected antenna temperature increase confirms that we have 
a good understanding of the antenna and RF system behavior, as well as of instrument orientation 
tracking. 

\subsection{Impulsive RF background survey}

\begin{figure}
\includegraphics[width=75mm]{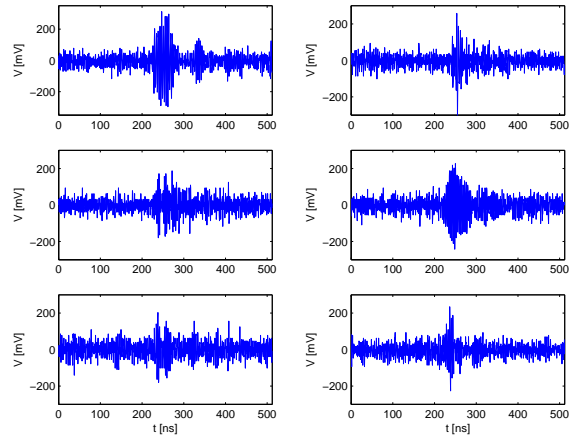}
\caption{\label{fig:bgevents}Examples of the six most common types of impulsive events recorded. Since 
they are repeatable and have no correlation with instrument location or orientation, we attribute them
to locally generated RF noise.}
\end{figure}

\begin{figure}
\includegraphics[width=65mm]{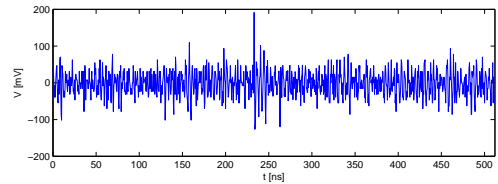}
\caption{\label{fig:signal}An example Askaryan pulse as would be seen by ANITA-lite.}
\end{figure}

The impulsive RF background analysis is based on 87,475 events recorded during 16.2 days of data taking
with 38\% average livetime. By comparing the recorded waveforms with the expected shape of 
coherent Cherenkov radio pulses, we were able to reject all of the recorded events while retaining 
62\% of simulated signal pulses. 
% Loosening the selection cuts and visually inspection and rejecting remaining few data events would
% boost the percentage of simulation signal passing the cuts.  
Over 90\% of all recorded events can be classified into six event
categories (Figure~\ref{fig:bgevents}), which we believe are all due to radio noise generated aboard
the instrument\footnote{As a piggyback instrument, we could not enforce as strict radio quiet 
conditions as would be required for this type of measurement. This will not be a problem with ANITA.}.
Figure~\ref{fig:signal} shows an example of the expected coherent Cherenkov radio pulse as would be
recorded by ANITA-lite.
Since we did not observe any radio pulses that could be associated with high-energy neutrino 
interactions in the ice, we can place a limit on the high-energy neutrino flux based on the ANITA-lite data. 
Currently, we are investigating the ANITA-lite trigger efficiency which is required to set the limit. 

\section{NEUTRINO FLUX SENSITIVITY}

\begin{figure}
\includegraphics[width=75mm]{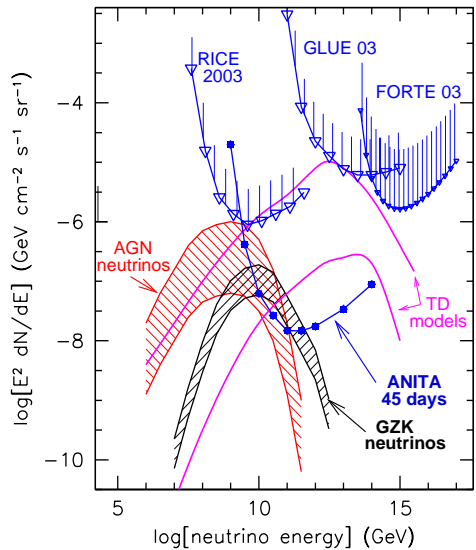}
\caption{\label{fig:limits}Published RICE, GLUE, and FORTE neutrino
flux limits, and the expected sensitivity of ANITA. The models of neutrino sources are given as 
bands, spanning the range from optimistic to conservative ones.}
\end{figure}
Simulating the performance of the full ANITA instrument and assuming an average flight trajectory
over Antarctica, we can calculate an expected flux sensitivity and compare it with published high-energy
neutrino flux limits and with models of neutrino flux (Figure~\ref{fig:limits}). In the figure, the 
sensitivity is given for 45 days of flight.  
An average long duration balloon flight above Antarctica is 15 days, although in 2004/05 season 
CREAM instrument flew for 42 days~\cite{cream}. 

\begin{figure}
\includegraphics[width=75mm]{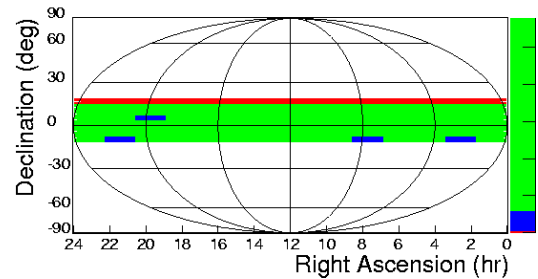}
\caption{\label{fig:skyview}Sensitivity of ANITA as function of neutrino arrival declination and 
right ascension. The green region indicates solid angle from which 90\% of the detectable flux originates, 
blue and green region account for 99\%, while blue, green and red account for 99.9\%. Neutrinos of 
$10^{19}$~eV were simulated.}
\end{figure}

ANITA will be sensitive to neutrinos arriving from low declinations (Figure~\ref{fig:skyview}). 
Instantaneous sky coverage is few degrees wide, but by circumnavigating Antarctica, the instrument
will observe the region indicated in the figure.

\section{CONCLUSIONS}

The Askaryan effect opens the possibility for use of huge quantities of Antarctic ice
as a neutrino detection medium. ANITA will be the first experiment with sufficient sensitivity to 
test current models of neutrino production due to GZK process and to probe for other astrophysical 
sources of neutrinos in the energy range from 0.1 to 100 EeV. ANITA-lite was a successful test of 
almost all subsystems planned for the ANITA instrument. We have established that Antarctica is
a very radio quiet environment, suitable for a search of neutrino-induced radio pulses. Also, we have 
demonstrated required timing precision and RF system gain calibration needed to perform accurate
measurements of such radio pulses.

\bigskip % extra skip inserted
\begin{acknowledgments}
The following organizations have provided support for this research: NASA Research Opportunities in Space Science (ROSS) -- The UH grant number is NASA NAG5-5387, Research Opportunities in Space Science -- DOE, Office of Science -- NSF and Raytheon Polar Services for Antarctic Support -- NSBF for Balloon Launching and Operations -- TIGER Collaboration for allowing ANITA-lite to fly as a piggyback.
\end{acknowledgments}

\bigskip % extra skip inserted
%\begin{thebibliography}{9}   % Use for  1-9  references

\end{document}